# A primary insight into the molecular phylogeny of *Colias* FABRICIUS, 1807 (Pieridae, Coliadinae) complex of South America

Alexander V. Kir'yanov

*Centro de Investigaciones en Optica*
*Loma del bosque 115, Col. Lomas del Campestre, Leon 37150, GTO., Mexico*

kiryanov@cio.mx

## Abstract

The data on molecular phylogeny of the *Colias* FABRICIUS complex of South America (SA), obtained via barcoding a mitochondrial part of genome, are reported. Barcoding was trialed employing the Barcoding of Life Database platform and comprised 93 specimens of SA *Colias* species and 2 outgroup specimens (*C. philodice guatemalena* RÖBER from North America). It is established that *Colias* species from SA form a single monophyletic clade, characterized by notable interspecific mutational divergences (3.5–4.5%) and moderate intraspecific ones (0.3–0.9%). It also reveals the occurrence of three well-established, or notably diverged genetically, species (*C. dimera* DOUBLEDAY & HEWITSON, *C. vauthierii* GUERIN-MENEVILLE, and *C. alticola* GODMANN & SALVIN stat. nov., the latter re-raised to specific level), plus a complex of three 'emerging' species (*C. lesbia* FABRICIUS, *C. euxanthe* FELDER & FELDER, and *C. flaveola* BLANCHARD), weakly differentiated in mitochondrial part of genome, with intersecting haplotypes. Furthermore, it disregards the specific status of taxa *weberbaueri* STRAND, *mossi* ROTHSCHILD, *nigerrima* FASSL, *erika* LAMAS, *blameyi* JÖRGENSEN, and *mendozina* BREYER, as certain evidences are being brought to consider them as subspecies of a clinal type of 'super-species' *C. flaveola s. lat*. The clade formed by all SA *Colias* is found to be sister to the clade that comprises all *Colias* species occurring outside the region (i.e. in Eurasia, North America, and Africa). A direct comparison of the data obtained for SA *Colias* with those for worldwide *Colias* available in GenBank data leads to the conclusion that these two parties are largely divergent autochthonous subcomplexes, being in fact subgenera of genus *Colias s. lat*. The first subgenus comprising all SA *Colias* species and named here *Aucolias* subgen. nov. (*Austral Colias*, or of Southern hemisphere stem) is sister to the second one comprising all *Colias* species from the Old and New Worlds, for which the name *Colias* (*Colias*) *s. str*. is preserved. The proposed subdivision of the genus is advocated by an insight into the historical development of the subgenera concept for *Colias* which reveals that the proposed subdivision deletes the inconsistencies that were inherent to the preceding attempts of *Colias* systematization.

## 1. INTRODUCTION

The group of *Colias* of South (S.) America is an interesting 'brick' of the genus, but so far, it has been poorly investigated. Although all the species (*sp.*) level entities that form the complex of S. American *Colias* were described in the 19th and 20th centuries, (see e.g. the summarizing works [VERHULST 2000 a-b, VERHULST 2013]), their *ssp.* are still and continuously being described (see e.g. [VERHULST 2016, KIR'YANOV 2017]). On the other hand, data on S. American *Colias* biology are very limited (apart from sporadic insights into their life-cycle, behavior, host plants, genitalia, seasonal, and geographical variability, etc., addressed in [MÜLLER 1957, PETERSEN 1963, SHAPIRO 1985, BERGER 1986, DESCIMON 1986, SHAPIRO 1989 (91), SHAPIRO 1992, SHAPIRO 1993, VERHULST 2000 a-b, LAMAS 2004, VERHULST 2013, GRIESHUBER ET AL. 2012, BENYAMINI ET AL. 2014]), and there are no reports neither on phylogenetic relations among *Colias* taxa within the range (S. America) nor on their relationship as a whole to *Colias sp.* occurring in other continents. Particularly, to the best of the author's knowledge, none has been reported so far on the molecular phylogeny of S. American *Colias*, though research works concerning *Colias* butterflies, inherent to some regions of the Northern Hemisphere, are progressively developed [WATT 1995, BRUNTON 1998, POLLOCK ET AL. 1998, WANG & PORTER 2004, WHEAT ET AL. 2005, BRABY ET AL. 2006, WHEAT & WATT 2008, LUKHTANOV ET AL. 2009, SCHOVILLE ET AL. 2011, WATT ET AL. 2013, LAIHO & GUNILLA 2013, WAHLBERG ET AL. 2014, DWYER ET AL. 2015, KRAMP ET AL. 2016, LIMERI & MOREHOUSE 2016 ]. However, the need for genetic studies that would clarify the phylogeny of *Colias* genus was prospected as early as the mid-20$^{th}$ century [HOVANITZ 1944, HOVANITZ 1957]. In the meantime, note that first studies on the molecular phylogeny of butterflies, characteristic to some areas of S. America, just got recently started (see e.g. [ELIAS ET AL. 2009, LAVINIA ET AL. 2017]).

For anybody aware of S. American *Colias* (especially, for those with fieldwork experience), it is nearly apparent that this group is different in many aspects from what is known about their relatives, inherent to Eurasia and North (N.) America. Though intuitive, this 'feeling' has influenced the author to undertake a study aiming, first, at gaining an insight into the molecular phylogeny of 'proper' S. American *Colias* as a concise complex and, second, at uncovering the type of its relationship to *Colias* of the rest of the World.

The method employed here to address the phylogenetic relations within *Colias* as genus is grounded on the concept of 'barcoding', where mitochondrial DNA (as a non-recombining maternally-inherited part of the genome) is under scope for treating mutational divergences of organisms at the intra- or inter-specific levels. Note that, in this method, a 648 base-pair (bp) region of mitochondrial DNA, corresponding to 5' segment of the cytochrome oxidase subunit I (CO I), is used as a marker, or 'DNA-barcode' [HEBERT ET AL. 2003 a-b]. Recently, barcoding has been framed within a resource, now known as 'BOLD' ('Barcoding for Life Database', Guelph, Ontario, Canada; see www.boldsystems.com). Its application allowed to get in the following years interesting data on barcoding of many insects, including *Colias* butterflies [LUKHTANOV ET AL. 2009, LAIHO & GUNILLA 2013, HUEMER ET AL. 2014, HEBERT ET AL. 2016, SIKES 2017].

In the meantime, some shortages and limitations are inherent to barcoding as a method of taxonomic and evolutionary studies; see e.g. [LEITE 2012] and the references therein. Sometimes, especially in questionable cases, a phylogenetic trial based on more versatile markers of the mitochondrial or/and nuclear part of genome is required for drawing an explicit pattern of the phylogenetic relations of a group of organisms under treatment.

In the present study, I report the data on the molecular phylogeny, resulting from a trial that was realized on the BOLD platform, over a set of *Colias* specimens stemming from the S. American continent. The first part of the results to report was obtained on the base of a console 'SACOL' ('South-American COLias'), while the second part was created after combination of the sequences generated within this console (for S. American *Colias*) with the ones, available in Genbank for some overseas *Colias* specimens.

Although the data obtained through barcoding (which handles a single fragment of the mitochondrial part of the genome) might seem to have limited value and, certainly, a more complex research, covering other regions of genome, would be required [POLLOCK ET AL. 1998, WHEAT ET AL. 2005, BRABY ET AL. 2006, LEITE 2012, WHEAT & WATT 2008, LAIHO & GUNILLA 2013, WATT ET AL. 2013, WAHLBERG, ET AL. 2014, KRAMP ET AL. 2016, MURILLO-RAMOS ET AL. 2016], fortunately, the current study reveals that whence applying this 'simple' routine, the generated phylogeny trees are quite stable and reproducible and the statistics of the calculated nearest-neighbor (further – 'NN') distances is confident, and so forth. This circumstance has been eventually defined by a type of the material under trial, viz. by the fact of the matter that *Colias* of S. America have turned out to be so genetically divergent (at the *sp.* level) that the data obtained leave one's little doubts.

## 2. MATERIALS

The individuals submitted for barcoding (in total 95, all belonging to the genus *Colias*) have been obtained in the wild in the Americas, with more than 95% of them collected by the author during the last 15 years. Of these, 93 individuals came from S. America (the territory ranging from Central Colombia to Southern Chile and Argentina), whilst the remaining two – chosen as an 'outgroup' reference, *C. philodice guatemalena* RÖBER, – came from southern N. America (S. Mexico, Chiapas) [EMMEL 1963]; see Table 1. The basic set of 93 *Colias* specimens represents almost all the *sp.* and *ssp.* of the genus that are currently recognized for S. America [VERHULST 2000 a-b, GRIESCHUBER AT AL. 2012]: *Colias dimera* DOUBLEDAY & HEWITSON, (nominotypical *sp.*); *Colias vauthierii* GUERIN-MENEVILLE, (*ssp. vauthierii* and *cunninghamii* BUTLER); *Colias lesbia* FABRICIUS (*ssp. lesbia, dinora* KIRBY, *andina* STAUDINGER, *verhulsti* BERGER, *mineira* ZIKÁN, and *misti* KIR'YANOV); *Colias flaveola* BLANCHARD (*ssp. flaveola, mossi* ROTHSCHILD, *mendozina* BREYER, *blameyi* JÖRGENSEN, *weberbaueri* STRAND, *nigerrima* FASSL, and *erika* LAMAS); *Colias euxanthe* FELDER & FELDER (*ssp. euxanthe, hermina* BUTLER, *coeneni* BERGER, *stuebeli* REISSINGER, and *alticola* GODMANN & SALVIN). Note that the chosen subdivision is made after [LAMAS 2004], but with warnings that: (i) Taxa *flaveola, mossi, mendozina, blameyi, weberbaueri, nigerrima,* and *erika* are considered herein as *ssp.* of *C. flaveola s. lat.*, though there exists an alternative viewpoint that all these, but excluding *nigerrima*, are separate *Colias sp.*, see e.g. [VERHULST 2013, BENYAMINI, et al. 2014]; the same (viz. a *sp.* status) is assumed by some authors concerning taxon *verhulsti*, but herein it is assumed to be a *ssp.* of *Colias lesbia*. (ii) *C. euxanthe*'s *ssp. euxanthe, hermina*, and *coeneni* are synonymized herein under the name *C. euxanthe euxanthe*, because of lack of any diagnostic characters that might unambiguously evidence the opposite. (iii) The *ssp. C. lesbia mineira* and recently described *ssp. C. flaveola benyaminii* [VERHULST 2016], *C. flaveola cora* and *C. flaveola laura* [KIR'YANOV 2017] were off scope here due to unavailability of material.

Specimens of these *Colias*, ranging at the *sp.* and *ssp.* levels (see Table 1), served as an experimental base for the SACOL console in BOLD. For each taxon, 2 to 8 specimens coming from each locality (except for *C. flaveola mendozina*, available as a single specimen) were taken for barcoding, with the idea being to gain insight into the statistical aspects of genetic divergences (NN distances), counted both within the genus (*Colias*) and within either *sp*. Note that similar actions have been taken, for comparison, at barcoding of overseas *Colias*, available in GenBank (where so far, however, there have been almost no data uploads for *Colias* from the S. American continent).

The whole of the *Colias* individuals that served as a source for the analytical work are preserved in the author's collection (Moscow, Russian Federation). Each butterfly providing a specimen under test was photographed, with the resulting images becoming available at the author's console SACOL in BOLD. Prior to making the analysis – comprised extracting DNA and their further processing at the Canadian Center for DNA barcoding (CCDB, Guelph, Ontario, Canada) – single legs of the animals (henceforth 'specimens') were carefully dissected from bodies and posed into single microcells, filled with 96% ethanol (30 mg). These cells formed a plate that was then delivered to the Center for passing through the standardized barcoding.

Table 1: SACOL project: taxa, their designations, numbers of sequenced specimens, localities, and host plants.

| *Colias* sp. / ssp. | Designation on the map (Fig. 4) | Number of specimens | Country of occurrence | Host plant |
|---|---|---|---|---|
| **C. dimera** DOUBLEDAY & HEWITSON | ★ | **8** | Columbia | *Trifolium sp.* (repens, dubium) |
| dimera DOUBLEDAY & HEWITSON | | 8 | Ecuador | |
| **C. vauthierii** GUERIN-MENEVILLE | ▽ | **14** | Chile | *Adesmia sp.* |
| vauthierii GUERIN-MENEVILLE | | 10 | Argentina | |
| cunninghamii BUTLER | △ | 4 | | |
| **C. lesbia** FABRICIUS | | **34** | Ecuador | *Medicago sativa* *Medicago sp.* *Trifolium sp.* *Astragalus sp.* |
| lesbia FABRICIUS | ✦ | 7 | Peru | |
| dinora KIRBY | ✦ | 6 | Bolivia | |
| andina STAUDINGER | ✦ | 5 | Chile | |
| verhulsti BERGER | ✦ | 12 | Argentina | |
| misti KIR'YANOV | ★ | 4 | Brasilia | |
| mineira ZIKÁN | | 0 | Uruguay | |
| **C. flaveola** BLANCHARD | | **27** | | *Astragalus sp.* *Medicago sp.* |
| flaveola BLANCHARD | ☐ | 3 | | |
| mossi ROTHSCHILD | ■ | 2 | | |
| mendozina BREYER | ■ | 1 | Peru | |
| blameyi JÖRGENSEN | ■ | 3 | Bolivia | |
| weberbaueri STRAND | ■ | 5 | Chile | |
| nigerrima FASSL | ■ | 3 | Argentina | |
| erika LAMAS | ☐ | 10 | | |
| benyaminii VERHULST | | 0 | | |
| cora KIR'YANOV | | 0 | | |
| laura KIR'YANOV | | 0 | | |
| **C. euxanthe** FELDER & FELDER | | **10** | Columbia | *Astragalus sp.*, *Trifolium sp.* |
| euxanthe FELDER & FELDER | ● | 2 | Ecuador | |
| = hermina BUTLER | ● | 2 | Peru | |
| = coeneni BERGER | ● | 2 | Bolivia | |
| stuebeli REISSINGER | ● | 2 | Chile | |
| alticola GODMANN & SALVIN | ● | 2 | | |
| **C. philodice** GODART | ✥ | **2** | Mexico | *Trifolium sp.* |
| guatemalena RÖBER | | 2 | Guatemala | |

The distribution of the specimens over the time elapsed after the capture events, in full accord with the collection members' labels, is presented in Fig. 1. As seen, the majority of these were less than 10 years in age (counted prior to the time of sequencing). In turn, the distribution of specimens over number for each taxon is provided in Table 1. Most of the specimens studied came from *sp. Colias flaveola / C. lesbia*, with the intention being an attempt to differentiate these two *sp.* at the intra- and inter-specific levels.

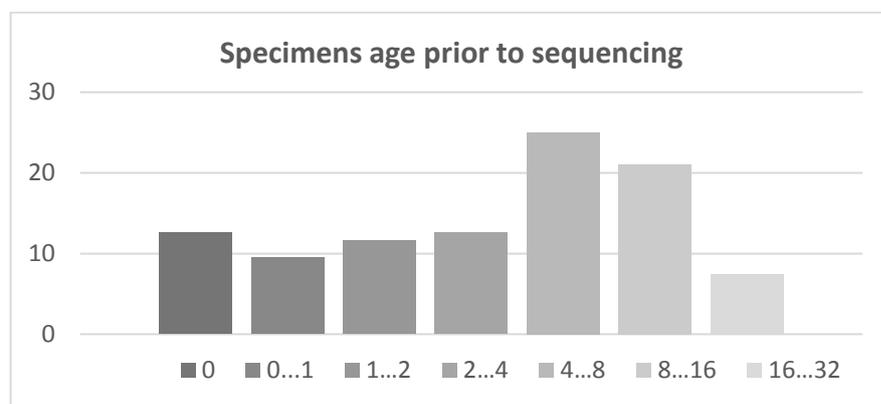

**Figure 1.** SACOL project: Distributions of specimens (as percentage of total number) over time elapsed from the butterflies' capture (counted prior to sequencing); each vertical bar shows a year-domain of counting.

## 3. METHODS

DNA extracts were prepared from a single leg of a specimen. On behalf of the author, DNA extractions, polymerase chain reaction (PCR) amplification, cycle sequencing and sequence analyses were performed applying the standardized protocols at CCDB. Sequences, electropherograms, and primers' details for all specimens forming the SACOL console were uploaded to BOLD; currently, these are being uploaded in GenBank for potential public use. The data were analyzed using the workbenches and tools, available on BOLD [Ratnasingham & Hebert 2007]. Note that all these were pre-analyzed to ensure their quality in terms of lack of contaminants and stop codons (the latter serves to insure a probabilistically negative role of 'pseudogenes', or nuclear mitochondrial DNA [Leite 2012]), but both were found to be null; neither sequence in the set has been determined as lacking unsuccessful traces. Furthermore, the absence of any subjective misidentification of the specimens was approved by BOLD's identification machine (all of them were pre-identified by the author prior to submitting to BOLD).

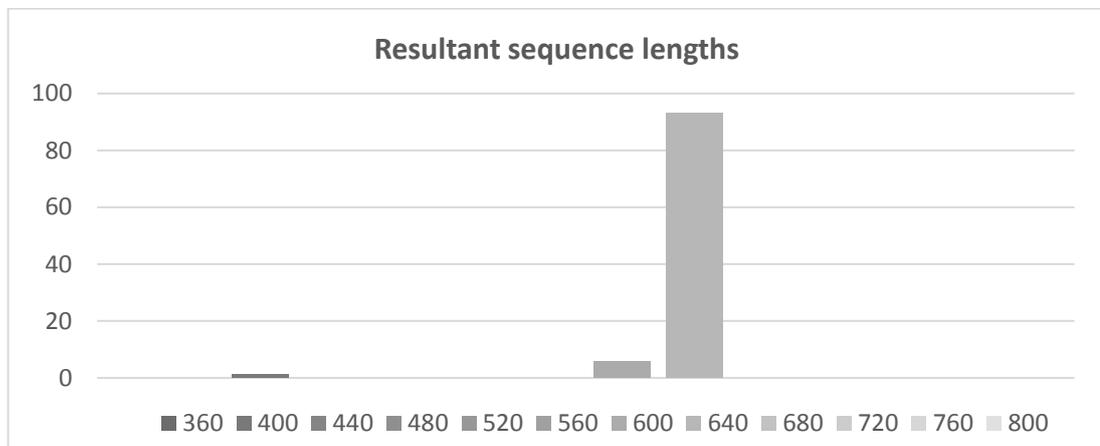

**Figure 2.** SACOL project: Distributions of specimens (as percentage of total number) over sequence lengths in bp of the CO I fragment under test (maximum – 648-bp length); each vertical bar corresponds to 40 bp.

Experimentally, sequence diversity was examined using a HMM (hidden Markov model) profile of a specific 648-bp fragment of the mitochondrial CO I gene (its 5' region). The DNA sequences were PCR-amplified using the primer pair (usually accepted for Lepidoptera) LEP-F1 (forward 5'-ATTCAACCAATCATAAAGATAT-3') and LEP-R1 (reverse 5'TAAACTTCTGGATGTCCAAAAA-3'). For details of cycling, processed via the standardized technique, see e.g. [Hebert et al. 2003 a-b, Ratnasingham & Hebert 2007, Hebert et al. 2016]. Specifically, the specimens were passed 2 to 5 times to ensure repeatability and reliability of the generated sequences; the data were assembled and analyzed using the BOLD Aligner (using a HMM based analysis). To generate neighbor trees in phylogenetics, cladistic, and unrooted versions (see the examples below), the standard Kimura-2-parameter distance model [Kimura 1981] was employed, without filtering. Note that all data presented below were generated from the sequences obtained at a limit of >500 bp, a criterium that provides, at worst, 1% uncertainty of base calls. As seen from Fig. 2, such DNA sequences were obtained from more than 99% source specimens, including those stemming from the older material. High confidence in the results obtained (viz., the generic trees and the distance summaries, see Section 4) is ensured by high quality and reliability of the input (viz. the experimental income itself and the generated sequences; see Fig. 3).

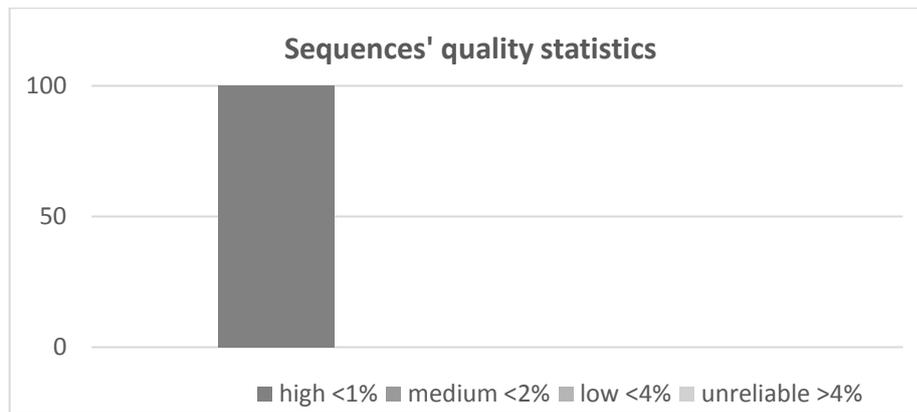

**Figure 3.** SACOL project: Distributions of specimens (as percentage of total number) in terms of sequence quality, ranged by degree of overall confidence (in turn, represented in uncertainty percentage).

Phylogeny trees were built either on the sequences obtained for all 93 S. American *Colias* specimens plus 2 outgroup *Colias philodice guatemalena*, forming the body of SACOL, or, additionally, on their representatives merged with the sequences obtained from *Colias sp.* of overseas provenance plus the ones sourced by 'neighboring' *sp.* of genera *Zerene* Hübner and *Aphrissa* Butler [Wheat et al. 2005, Braby et al. 2006, Wheat & Watt 2008], available in GenBank. In the last case, correctness and easiness of processing was provided by full compatibility of these, given the same analytical resources, standardized at the BOLD platform, employed.

Note that the statistical pairwise analyses of the data were feasible directly on the BOLD platform. The results of such analytical tests are highlighted below in the form of tables where average mutational distances between the correspondent haplotypes (intra-*sp*. divergences) are compared with the ones within haplotypes (inter-*sp*. divergences). The latter data seem to be valuable for making estimates for both internal genetic divergence of S. American *Colias* and as the whole relatively to their overseas sisters and for revealing the types of intra-group divergences, inherent to *Colias* of S. America overall versus to outgroup *Colias sp.* (of the rest of the World).

## 4. RESULTS AND DISCUSSION

The mapping of the material that served as source for the SACOL project is presented in Fig. 4, with collecting places marked by symbols of different kinds that identify each *Colias sp./ssp*. (different colors highlight the ground coloration of males' verso). Note that the explicit information about the barcoded specimens, as well as the sequences generated, are available on BOLD (SACOL console: specimens AVK-001 to AVK-095).

The phylogeny trees generated as the output of barcoding are demonstrated in Figs. 5 to 9. Figs. 5 – 7 provide a resume of the results for S. American *Colias*, framed by SACOL only whilst Figs. 8 – 9 present an extended view on branching pattern of *Colias* as genus (discussed in more detail in Section 5). In the last case, the data from sequencing SACOL's specimens are combined with those for worldwide *Colias*, available in GenBank. In Fig. 7 are provided the illustrations of some of the adults from which the specimens under study came to facilitate one's understanding of the relationship between the butterflies' phenotypes and their generic positions on the trees generated. Fig. 10 shows the statistical 'fingerprints' (in terms of NN distances) of S. American *Colias* (SACOL's data) against overseas *Colias sp.*, stemming from N. America, Eurasia, and Africa

(GenBank's data), as the result of barcoding. Tables 1 to 3 permit one to elucidate some of the details, discussed in the text.

In Fig. 4, the localities of capture of all 95 *Colias* butterflies, establishing the whole barcoded specimens (SACOL) are mapped. For easiness of attributing the material, the symbols designating each *sp./ssp.* in Fig. 4 (as these were a priori determined by the author) are copied to Table 1. As seen from Fig. 4 and Table 1, the collecting range covers the whole S. American continent, from the north (Colombia) to the south (S. Patagonia), while it is mainly restricted to the Andes and adjacent territories, where abundance of *Colias* taxa is the highest (as they mostly prefer temperate environments where their host plants grow in suitable numbers).

The places where the specimens of the taxa under scope of this study (*Colias dimera*; *Colias vauthierii* (*ssp. vauthierii* and *cunninghamii*); *Colias lesbia* (*ssp. lesbia, dinora, andina, verhulsti,* and *misti*); *Colias flaveola* (*ssp. flaveola, mossi, mendozina, blameyi, weberbaueri, nigerrima, erika*); *Colias euxanthe* (*ssp. euxanthe = hermina = coeneni, stuebeli,* and *alticola*)) stemmed from are symbolized in Fig. 4. To complete the picture, the locality for outgroup *Colias philodice guatemalena* (S. Mexico) is added: see the map's upper left corner. Note that the points of type localities for the taxa belonging to S. American *Colias* are not marked on the map, but the reader may refer to [BERGER 1986, VERHULST 2000 a-b, VERHULST 2013] to make sure that the collecting places for the SACOL's specimens match them well.

Besides, I found it reasonable to provide on the map the place (highlighted as '*?*', the Magellan region) of probable provenance of the legendary *Colias ponteni* WALLENGREN = *C. imperialis* BUTLER, a butterfly never again collected after the capture of its type series and descriptions in the mid-19[th] century [SHAPIRO 1989 (91), SHAPIRO 1993]. The type series of this *Colias* were inaccessible for barcoding; so, this important representative of the *Colias* genus, as the most primitive amongst the others [PETERSEN 1963], was out of scope here.

In Fig. 5, I present the phylogenic tree obtained after barcoding the SACOL's specimens, i.e. all S. American *Colias* plus outgroup *guatemalena*. Note that in this tree, as well as in the ones shown below, horizontal distance between specimens' positions is a measure (in percentage) of relative mutational divergence. Accordingly, if a few specimens are grouping in clusters (branches), horizontal distance becomes a measure of mutational divergence between such clusters (being the whole, or a part of, subgenus, *sp.*, *ssp.*, 'lineage', or haplotype).

As seen from Fig. 5, all SACOL specimens are split into five well-separated and largely distinct clusters, distancing by mutational divergences of ~4.5% in average.

The two outgroup specimens of *C. philodice guatemalena* are joined together, forming the first cluster ('1'), significantly spaced off (by ~5.6%) from the rest of the specimens coming from 'proper' S. America); see also Section 5.

The presence of two other concise clusters ('2' and '3' in Fig. 5), one comprising all the *Colias dimera* specimens and the other all the *Colias vauthierii* specimens, perfectly supports the *sp.* status of the two entities. In the meantime, whereas the specimens of *C. dimera* are weakly differentiated within own cluster (evidencing nominotypical essence of this *sp.*), the ones of *C. vauthierii* are split into two well-differentiated sub-clusters (spaced by ~1%) that entirely correspond to the phenotypically different *ssp. vauthierii* and *cunninghamii*.

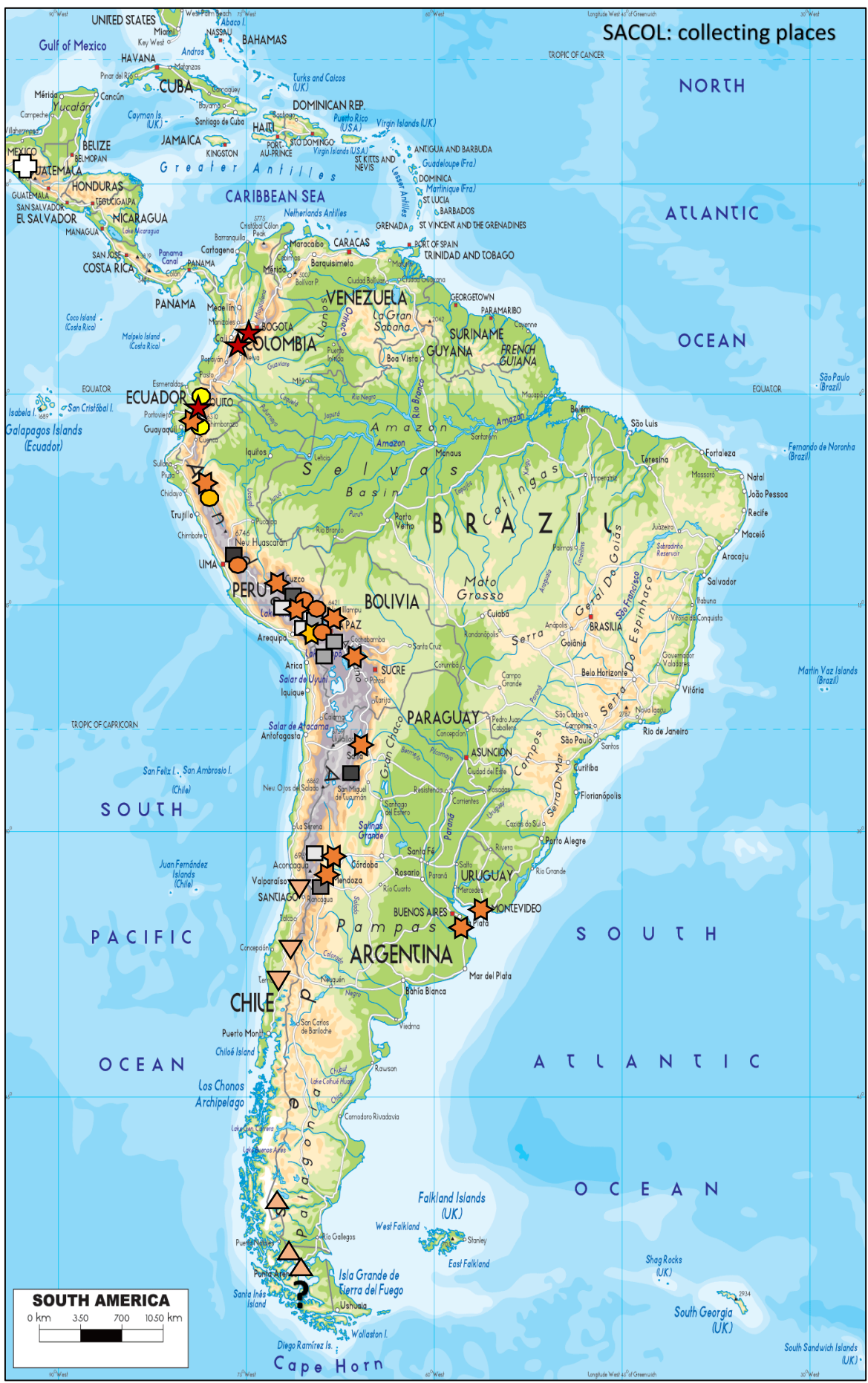

**Figure 4.** Mapping of collecting places (SACOL console). For details, see Table 1, Fig. 7, and the text.

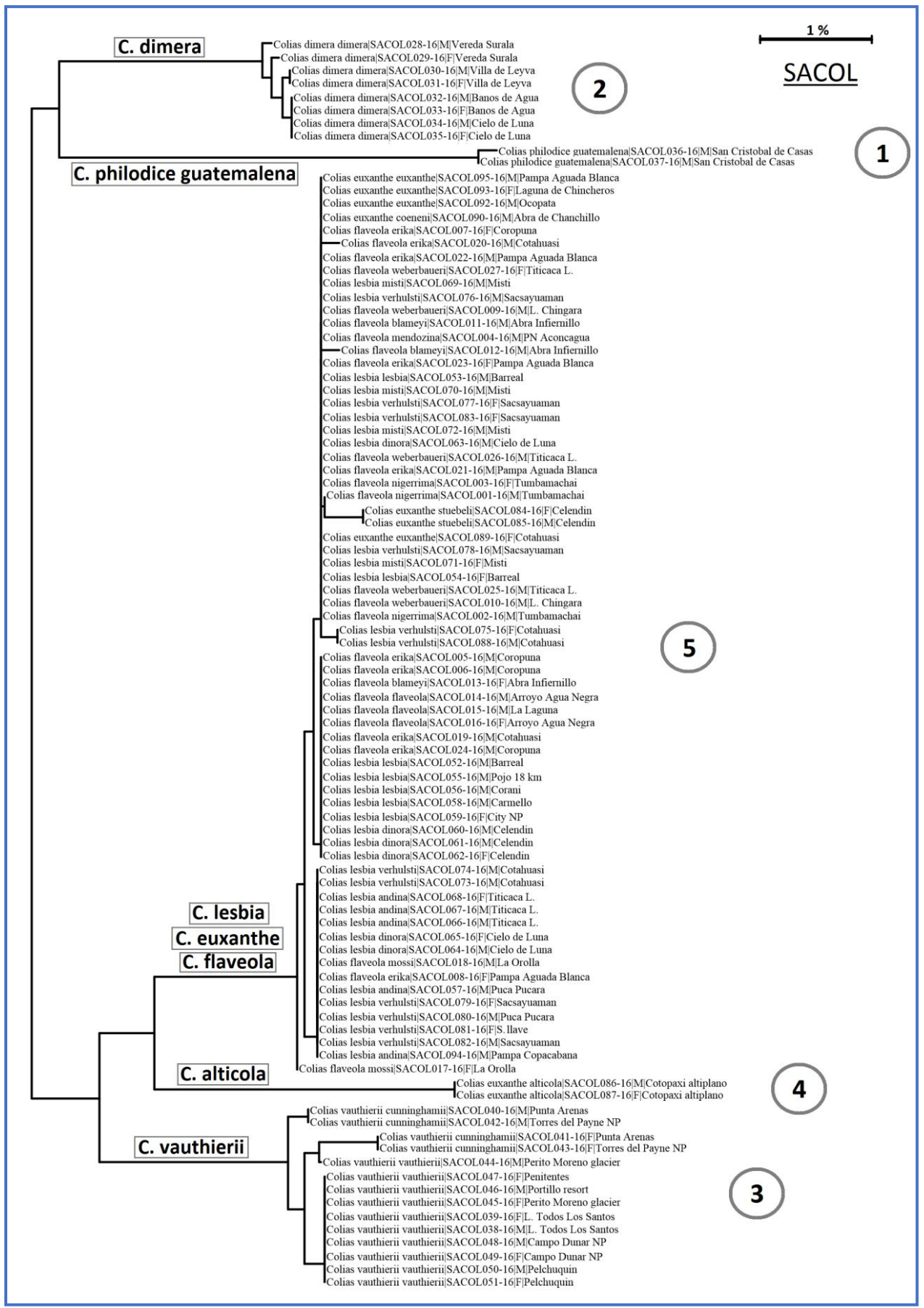

**Figure 5.** Hierarchical phylogeny tree (branching in relative distances), generated via 'barcoding' *Colias* specimens (SACOL console).

The situation with the rest of S. American *Colias (C. lesbia, C. flaveola,* and *C. euxanthe)* is more sophisticated and even controversial.

First, in Fig. 5, it is obvious that the two specimens of *C. euxanthe alticola* are 'cut off' from all other specimens of *C. euxanthe* (*ssp. euxanthe* and *stuebeli*) by >4% mutational divergence distance, thus establishing its own cluster ('4') in the tree. This casts no doubts that this taxon is in fact a separate (*bona*) *sp.*, hereafter referred to as *C. alticola* (stat. nov.). It is worth noting that *C. alticola* had been originally described as a distinct *sp.* with the same name [GODMANN & SALVIN 1891]. Hence, the action performed above is nothing more than the taxon's re-raising to its original rank.

Second, the remnant *ssp.* of *C. euxanthe*, taken together with all ssp. of *C. lesbia* and *C. flaveola*, are almost indistinguishable at a glance within cluster '5', comprising the majority of SACOL's specimens. This circumstance is surprising, given that, phenotypically, these three *sp.* are easily differentiable by the eye (Fig. 7), and, besides, they commonly have different plants as hosts (Table 1) and somewhere occur in sympatry. However, genetically, as seen from Fig. 5, these *sp.* deviate from each other slightly, by no more than 0.3%. Besides, the intraspecific divergences within each of these taxa – at the *ssp.* level – are quite comparable, ~0.15–0.2% (see below). Thus, primary overview of the genetic differences among *C. lesbia, C. flaveola,* and *C. euxanthe* (but excluding *C. alticola*) leads me to the conclusion that they constitute a complex of weakly deviated taxa of presumably young age, thus pointing their status as 'emerging' *sp.* Interestingly, insight into the fine structure of this complex shows that the haplotypes formed in the fifth cluster (with the tolerance of barcoding) are sometimes shared by these three *sp.* That is, some of the individuals, apparently belonging to one of them, according to their well-established phenotypes, are frequently 'scattered' across the haplotypes, which indicates close relatedness of *C. lesbia, C. flaveola,* and *C. euxanthe*.

Furthermore, it is worth noticing that butterflies of these three *sp.* (cluster '5'), when spatially overlapping in Nature – which happens frequently, given by their generally sympatric occurrence, – not rarely tend to copulate, with a variety of 'putative' hybrids produced, a fact known to people familiar with the biology of S. American *Colias*.

A few examples of such natural hybridization, figured in Table 2, may serve as support for the entities' genetical neighborhood and close relatedness, and hence, point at introgression and intensive gene flow amongst. In this sense, the presence of stable white-male forms in *C. euxanthe euxanthe* (viz. *dimorpha* HEMMING aberration), almost coinciding in habitus with white males of *C. flaveola erika* (see panel (d) in the Table), is of mention, too.

On the contrary, there are no data (and the author has never met such examples, either) about hybridizing in the wild between, say, *C. dimera* and *C. alticola* / *C. lesbia dinora* (in Ecuador) or between *C. vauthierii* and *C. lesbia* (in Central Argentina), even in the areas of their close neighborhood or sympatry (though such co-occurrence is rare).

The examples presented to illustrate hybridization among S. American *C. lesbia, C. flaveola,* and *C. euxanthe* (a natural phenomenon but correlating with the sharing of the entities' haplotypes, as established by the phylogenetic pattern shown in Fig. 5) recall a many-times visited case of hybridization among N. American *Colias*, e.g. between the genetically close *C. eurytheme* BOISDUVAL, *C. philodice* GODART, and *C. eriphyle* W. EDWARDS [GEROULD 1946, AE 1959, HOVANITZ 1963, WATT 1995, WANG & PORTER 2004, PORTER & LEVIN 2007, WHEAT & WATT 2008, ESTOUP & GUILLEMAUD 2010, JAHNER ET AL. 2012, SHAPIRO 2012, WATT ET AL. 2013, DWYER ET AL. 2015]. Note that in the last research, the authors discuss in detail genetic drivers of the phenomenon using the approach of [WANG & PORTER 2004] but conclude on incomplete resolving of the case. Besides, for Eurasia, a well-known example of genetically indistinguishable, on one side, but easily hybridizing, on the other, are *C. erate* ESPER and *C. croceus* GEOFFROY; see e.g. [LUKHTANOV ET AL. 2009].

**Table 2:** Examples of strong aberrant forms (most probably, 'putative' hybrids) in the complex *C. euxanthe / C. lesbia / C. flaveola* (a-b-c, e-f-g-h) and white-male form of *C. e. euxanthe* (d); top – males; bottom – females (the asterisked form (h) has passed barcoding, **\***). Compare the hybrid male-forms with common appearances of males in these *sp.* in Fig. 7.

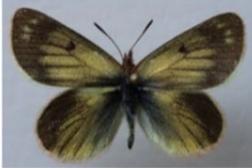

Concerning S. American *Colias* (*C. lesbia*, *C. flaveola*, and *C. euxanthe*), the interrelations of a tendency to hybridize, giving rise to the makeups exemplified in Table 2, and a high degree of sharing of haplotypes (refer to Fig. 5) by these phenotypically different *sp.* seem to be an interesting issue for further studies. In this sense, the questions posed in e.g. [LUKHTANOV ET AL. 2009] deserve attention: whether both effects are explained by recent speciation of these taxa (viz. their current evolutionary stage as 'emerging' *sp.*), or by a balanced character of sympatric occurrence (introgression against selection), in the recent past or presently, as genetically weakly diverged entities, having developed before as allopatric. However, one should account for the fact that phylogenetic identity of these 3 *sp.* is established here as the result of barcoding (where, remind, a sole fragment of mitochondrial DNA was under treatment). Hence, it cannot be ruled out that the mutations that are impactful for the natural separation of these three *sp.* might arise in very few different loci outside this fragment and so unrecoverable by the method used.

It is worth emphasizing that the phylogenic clusters (branches '3' to '5'), embracing *sp. C. vauthierii, C. alticola, C. lesbia, C. flaveola*, and *C. euxanthe*, form a single clade, or lineage, separated from the one of *C. dimera* of Northern S. America; furthermore, this group of *sp.* as the whole is largely spaced genetically from *C. philodice guatemalena* of southern N. America. This allows plotting the results of barcoding S. American *Colias*' in the form of an 'unrooted' tree: see Fig. 6. Note that trees of such type are helpful in addressing the fine structure of a group under scope (see e.g. [WATT ET AL. 2013, DWYER ET AL. 2015]) and, eventually, allowing to define – by means of estimating relative NN distances – a type of components (genera, separate *sp.*, *ssp.*, lineages, populations, etc.) that such a group is composed of, if average relative mutational divergences between evolutionary units of similar groups of organisms of this or genetically adjacent genera are known (or can be reliably estimated) from other sources.

From Fig. 6, on the one hand, one sees, again, the large genetic divergences between the 'good' (i.e. well established after long-time of separate historical development) *sp.*, viz. *C. dimera, C. vauthierii, C. alticola*, plus the group of 'emerging' *sp. C. lesbia, C. flaveola*, and *C. euxanthe*, as these four are spaced out by ~4 to ~5% changes in the mitochondrial DNA. On the other hand, all members of the latter complex (tentatively attributed here as *ssp.* of either of the 'emerging' *sp.*) are split by very small genetic divergences.

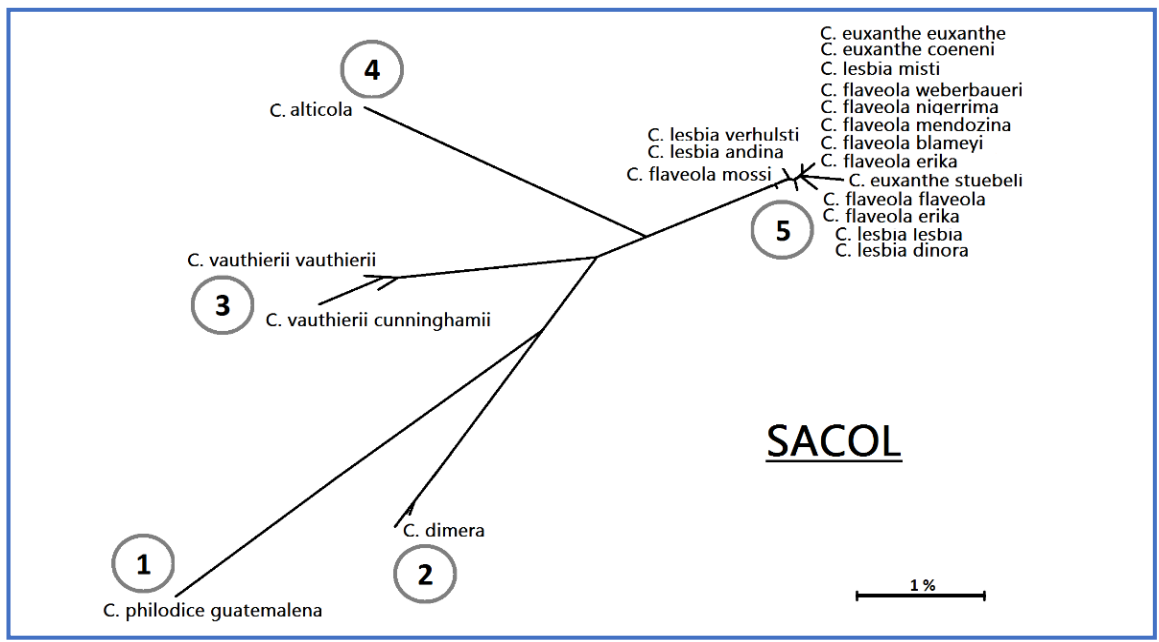

**Figure 6.** Unrooted phylogeny tree (for relative distances), generated applying barcoding of SACOL's *Colias* specimens.

Fig. 7 provides a summary of barcoding, focused on intra-specific (at the *ssp.* or populational level) mutational divergences of all S. American *Colias*, framed by SACOL console; it is illustrated by the images of some of the adults, received DNA sequencing.

As seen, all these (including *C. alticola*) are easily identifiable (and separable) by the eye, regardless of fine differences of *ssp.* phenotypes (i.e. deviations in ground coloration, wing shape, size, etc.). Noticeably, the visual perception correlates well with the data of barcoding, proceeded separately for each *sp.*

For instance, the adults of nominative *ssp.* and *ssp. cunninghamii* of *C. vauthierii*, as well as nominative *ssp.* and *ssp. stuebeli* of *C. euxanthe*, are seen to be phenotypically quite distinct; in turn, the genetic divergences between these are considerable, supporting the visual sorting: in the first pair, it is measured by ~0.9%, whereas in the second by ~0.4% (see the two upper panels in Fig. 7).

Of another kind is the circumstance of 'emerging' *sp. C. lesbia*, *C. flaveola*, and of *C. euxanthe*: see the panels in the middle part and bottom of Fig. 7.

For these *sp.*, both intra-specific and inter-populational divergences are notably small (~0.1–0.2%); this result is a bit surprising, because, say, the members (recognized *ssp.*) of *C. lesbia* complex are peculiar in overall habitus and ground coloration (e.g. see in the figure dirty-yellowish *C. lesbia misti*). However, notice the inconsistency between intra-specific phenotypic variability (mostly expressed in ground coloration) among *ssp.* of *C. flaveola*, 'fleeting' from snow-white in *ssp. flaveola* and *erika* to almost black in *ssp. nigerrima* and *mossi*, on the one side, and, on the other, their remarkable genetic homogeneity (~0.05% divergence); see the right-bottom panel in Fig. 7.

Finally, note that the case of *C. alticola* was addressed above; in fact, this taxon presents an infrequent example of hidden, or cryptic, *sp.* in the *Colias* genus.

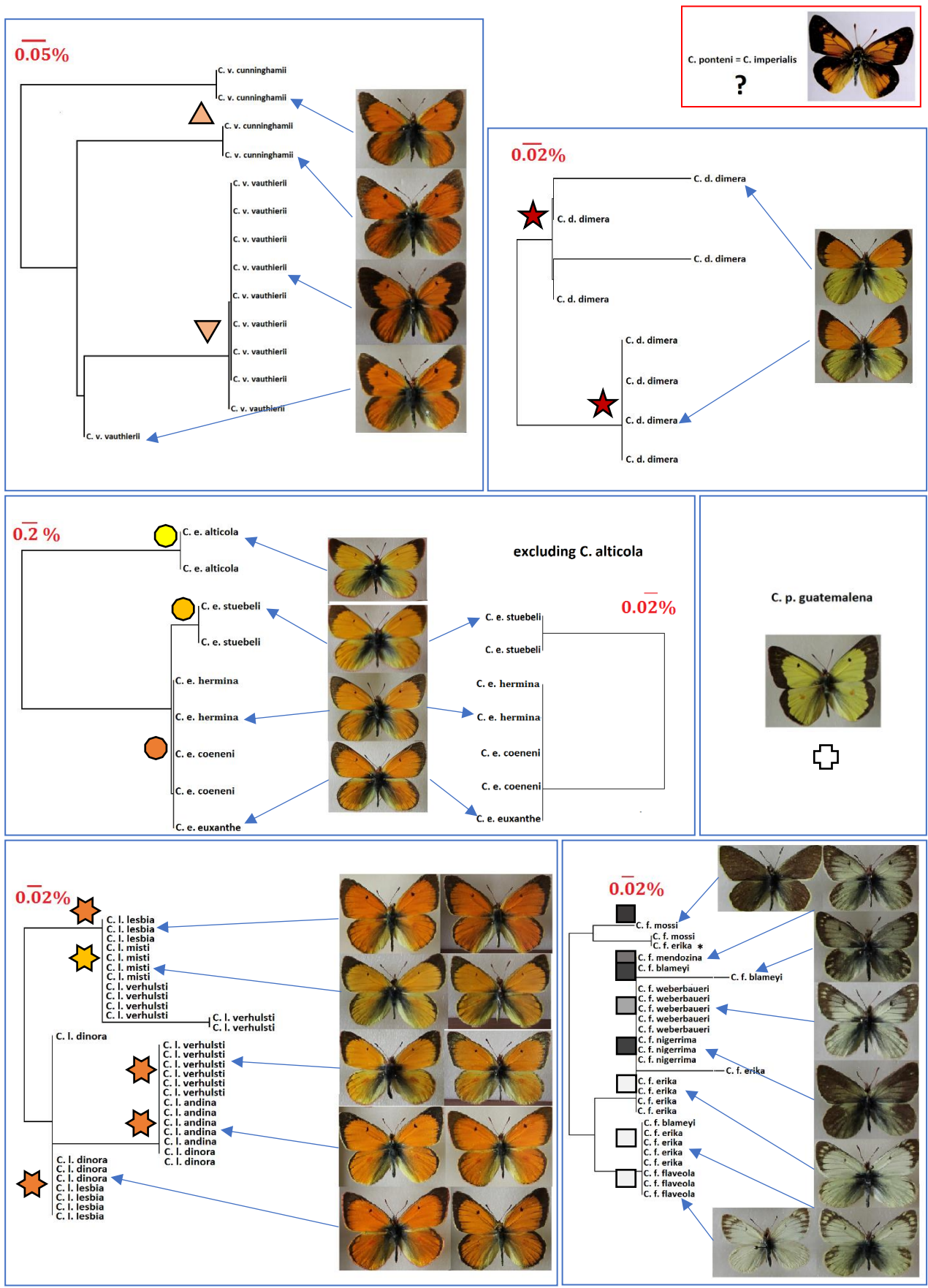

**Figure 7.** Subpanels on the left: phylogenic branches (haplotypes) of the tree plotted in Fig. 5, each generated for each *Colias sp.* covered by SACOL project, revealing their intra-specific divergence; symbols correspond to the legends used for mapping the taxa in Fig. 4. Distance scaling is always provided nearby. Subpanels on the right: images exemplifying the adults sourced for the barcoded specimens. On the upper right corner, the image of one of the type specimens of *C. ponteni = C. imperialis* is shown for comparison (the photo was taken on permission of the authorities of the British Museum of Natural History (London, UK).

All these trends, established after analyzing Figs. 5–7, are summarized in Table 3.

**Table 3:** Parameters of intra- and inter-specific relatedness in S. American *Colias*.

| sp. | Mean deviations of NN distances within taxon, % | Maximal deviations of NN distances within taxon, % | Average distances to NN within SACOL project's members (between sp.), % | Notes |
|---|---|---|---|---|
| *C. dimera* | 0.16 | 0.31 | 4.25 to *vauthierii* | homogeneous *sp.*: no apparent *ssp.*; variability on populational level only |
| *C. vauthierii (including ssp. cunninghamii)* | 0.38 | 0.93 | 3.52 to complex *lesbia-euxanthe-flaveola* as whole | two well-differentiable *ssp.* |
| *C. vauthierii (excluding ssp. cunninghamii)* | 0 | 0 | – | notably homogeneous *ssp.* |
| *C. vauthierii cunninghamii* | 0.56 | 0.93 | – | quite inhomogeneous in populations *ssp.* |
| *C. flaveola (all ssp.)* | 0.13 | 0.49 | <0.25 to *euxanthe* (but excluding *alticola*) and to *lesbia* | genetically hardly differentiated from *lesbia* & *euxanthe*; has many 'phenotypical' *ssp.* |
| *C. euxanthe (all ssp. including ex-ssp. C. euxanthe alticola)* | 1.77 | 4.73 | <0.25 to *lesbia* & *flaveola* | genetically hardly differentiated from *flaveola* & *lesbia*; [note: *alticola* is implied to be separate *sp.*] |
| *C. euxanthe (all ssp. excluding ex-ssp. C. euxanthe alticola)* | 0.19 | 0.46 | 3.56 to *alticola* | composed of two well-defined *ssp.* |
| *C. alticola* | 0 | 0 | 3.56 to all *ssp.* of *euxanthe* (but excluding *alticola*) | genetically & phenotypically is separate *sp.* |
| *C. lesbia (all ssp.)* | 0.17 | 0.49 | <0.25 to all *ssp.* of *euxanthe* (but excluding *alticola*) and to *flaveola* | genetically hardly differentiated from *flaveola* & *euxanthe*; has many 'topotypical' (genetically weakly diverging) *ssp.* |
| *C. philodice guatemalena* | 0.16 | 0.16 | 5.57 to *dimera* | southernmost *ssp.* of *sp.* belonging to a different *Colias* subgenus |

The data presented in the Table may be compared with the ones known for average divergences found in the literature for other (aside S. America) representatives of genus *Colias* [WHEAT ET AL. 2008]. For instance, for closely-related and cryptic *sp.* within the genus, genetic divergences are estimated to be 0.67–1.53%, whereas those for 'solid', or strongly differing in habitus (including the ones belonging to different *Colias* subgenera), *sp.*, these are estimated to be 0.55–2.26%; meanwhile, the divergences detected on sub- and semi-specific or inter-regional levels for *Colias* do not exceed 0.77%. Upon comparing this set of values with the values presented in Table 3, it becomes clear that, apart from 'emerging' *sp. C. lesbia*, *C. euxanthe*, and *C. flaveola* (almost indistinguishable in barcodes), the rest of S. American *Colias* demonstrate larger or much larger relative divergences: 3.5–4.3%; note that this law is particularly obeyed in the case of the complex of *C. lesbia*, *C. euxanthe*, and *C. flaveola* if it is treated as a single concise unit.

In the meantime, the distance of all S. American *Colias* to outgroup *C. philodice guatemalena* is notably larger (~5.6%): this value is comparable with the distance of the *Colias* genus as a whole from its sister, the genus *Zerene*. This points at the following hypotheses as least: (*i*) a history of S. American *Colias* is relatively older than that of overseas *Colias* and (*ii*) strong genetic diversification within the complex of well-differentiated sp. of S. American *Colias*, indicating independent evolutions for long time and without serious gene flow amongst.

More details of unusual appearance of S. American *Colias* against their relatives, occurring in the rest of the World, can be uncovered via straightforward comparison of these two subcomplexes;

the results are highlighted in Figs. 8–10. Here, the barcoding data of (*i*) the representatives of S. American *Colias* framed by the SACOL console and (*ii*) the ones of most overseas *Colias sp.* (from Eurasia, N. America, and N. Africa), sequences of which stemmed from GenBank, are provided. The combined set of the latter data has passed analytical processing, analogous to the one applied when handling solely the SACOL specimens (as described in Section 3). To ensure reliability of the trees generated, a sequence available in GenBank for *C. lesbia* (the only data for a *sp.* belonging to S. American *Colias* in the database) was in this case incorporated *ab initio* as input. This fact, together with the fact that sequences for *C. philodice guatemalena* (the reference outgroup *Colias* in the SACOL console) had many sequences in GenBank, received from its close relatives in N. America, *C. philodice* and *C. eurytheme*, provided a good base for an independent cross-check of the results.

The type of the hierarchical phylogeny tree shown in Fig. 8 permits to reveal the following laws. Note that the *Colias* subgenera, after [BERGER 1986], 'wrapped' as legends inside the tree plotted in Fig. 8, are discussed in Section 6, where the issue of *Colias* genus' subdivision is addressed.

First, all S. American *Colias* (including *C. dimera*) form a separate segment of the tree, largely divergent from *Colias sp.* inhabiting other continents, by more than 4.5% of relative distance. Second, the internal structure of the subcomplex comprising all S. American *Colias* (viz. *sp. C. vauthierii*, *C. alticola*, *C. lesbia*, *C. euxanthe*, *C. flaveola*, and *C. dimera*) is not altered by biasing the 'outside' worldwide *Colias sp.*, i.e. no 'cross-talks' between the former and the latter subcomplexes were ever generated. Third, the last subcomplex, comprising all *Colias* of the Old and New (but excluding S. American *Colias*) Worlds, appears to be generally of quite different, more polyphyletic, essence. Fourth, once considered as the whole, this subcomplex clearly becomes sister to the one comprising all S. American *Colias* (with *C. dimera* lineage of Northern S. America being its nearest relative). Fifth (in turn), for these two subcomplexes, the subgenera *stat.* can be of worthiness only (but see Section 5 for a more comprehensive treatment, though). Sixth, a detail that may validate reliability and stability of the tree built is that the reference elements chosen to enter the combined data set – *C. philodice guatemalena* from the SACOL console and *C. lesbia* from GenBank – are 'properly' nested by barcoding on the phylogeny tree; that is, they occupy the places explicitly adherent to their closest relatives – *C. philodice* and *C. eurytheme* of GenBank and *C. lesbia* (cluster '5', SACOL), correspondingly.

In Fig. 9, I show the unrooted phylogeny tree, built on some of the characteristic elements that constitute the hierarchical tree (Fig. 8), as an illustration that helps one to snapshot the basic novelties that the present study brings in. Namely, in Fig. 9 are delimited all the lineages formed by the subcomplex of *Colias* of S. America (ranged in terms of clusters '2' to '5' in Fig. 5), plus a few lineages of the subcomplex of overseas *Colias*: the one, represented by the reference *sp. C. philodice* / *C. eurytheme* and the other, composed of *C. hyale* (LINNAEUS) & *C. alfacariences* RIBBE, closest in NN distances to S. American *Colias*. The 'neighboring' genera, *Aphrissa* and *Zerene*, being genetic sisters to *Colias* [WHEAT ET AL. 2005], are mapped for a comprehensive comparison.

The dashed ellipses in Fig. 9 topologically specify single branches, or clades, of the tree, the first covering all S. American *Colias sp.* (thus, advocating its 'proper' status) and the second covering the 'remnant' (worldwide) *Colias sp.* The first clade – I believe I act unambiguously – is named hereafter '*Aucolias*', viz. '*Austral Colias*', highlighting the region of S. America as provenance, or 'the Southern Hemisphere', i.e. '*Austral America*'. In fact, it should be considered as a well-supported genetically new subgenus of the genus *Colias s. lat.*, having been detached at its primary branching from a hypothetical ancestor and subsequently split into own lineages.

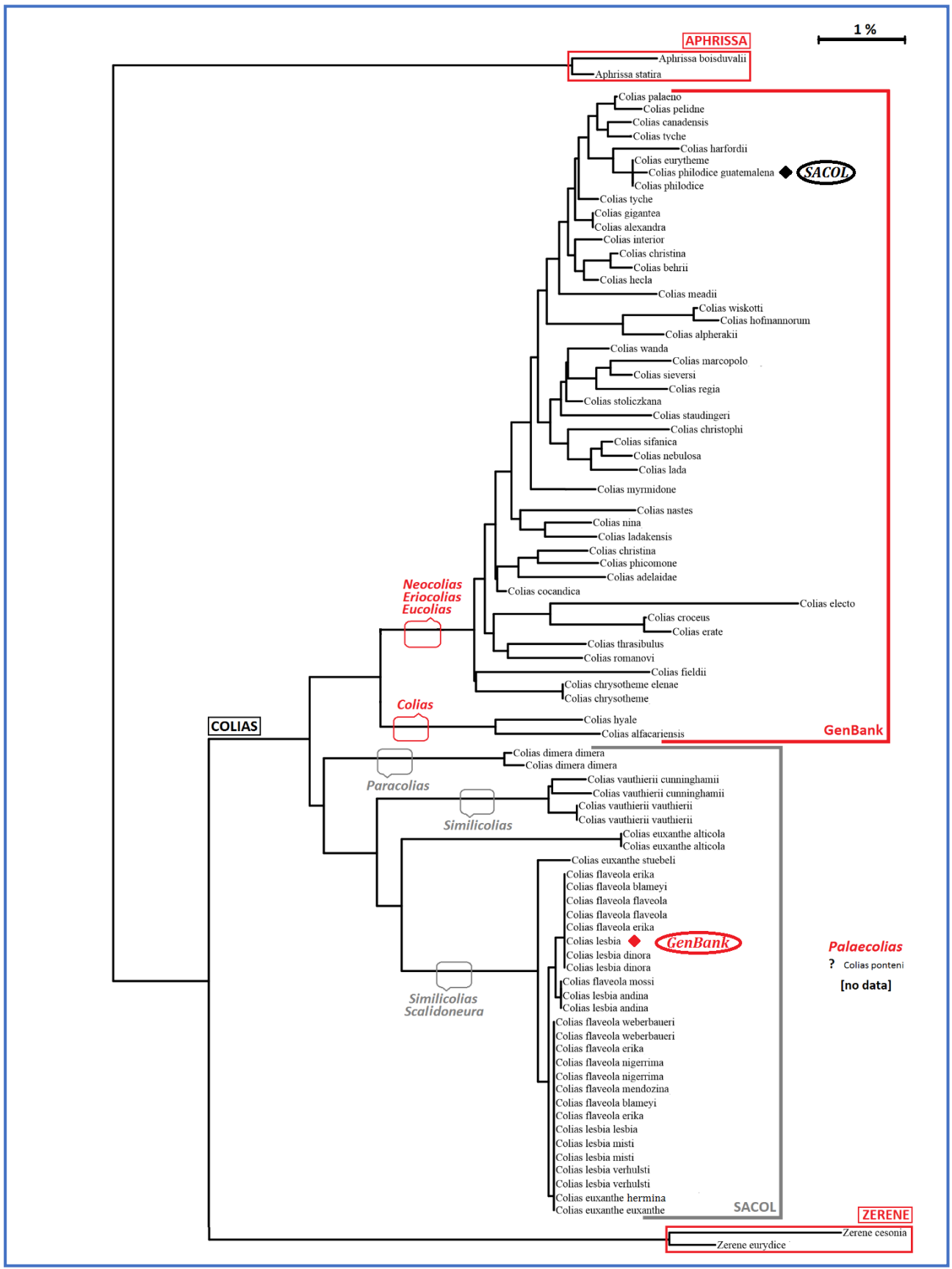

**Figure 8.** Hierarchical phylogeny tree (for relative distances), generated applying sequencing, built on the basis of the combination of the data stemming from 'barcoding' S. American *Colias* (SACOL project: highlighted in grey) and the ones obtained using an analogous routine from GenBank, sourced from characteristic representatives (*sp*.) of genus *Colias* from outside S. America, as well as from representatives (*sp*.) of genetically closest genera *Zerene* (sister to genus *Colias*) and *Aphrissa* (sister to genera *Zerene* / *Colias*); the whole of the data stemming from GenBank are highlighted in red. Warning: in italics (see also dashed boxes) is indicated the currently used subdivision on 8 subgenera of *Colias* genus [BERGER 1986], apparently inconsistent with the data of molecular phylogeny.

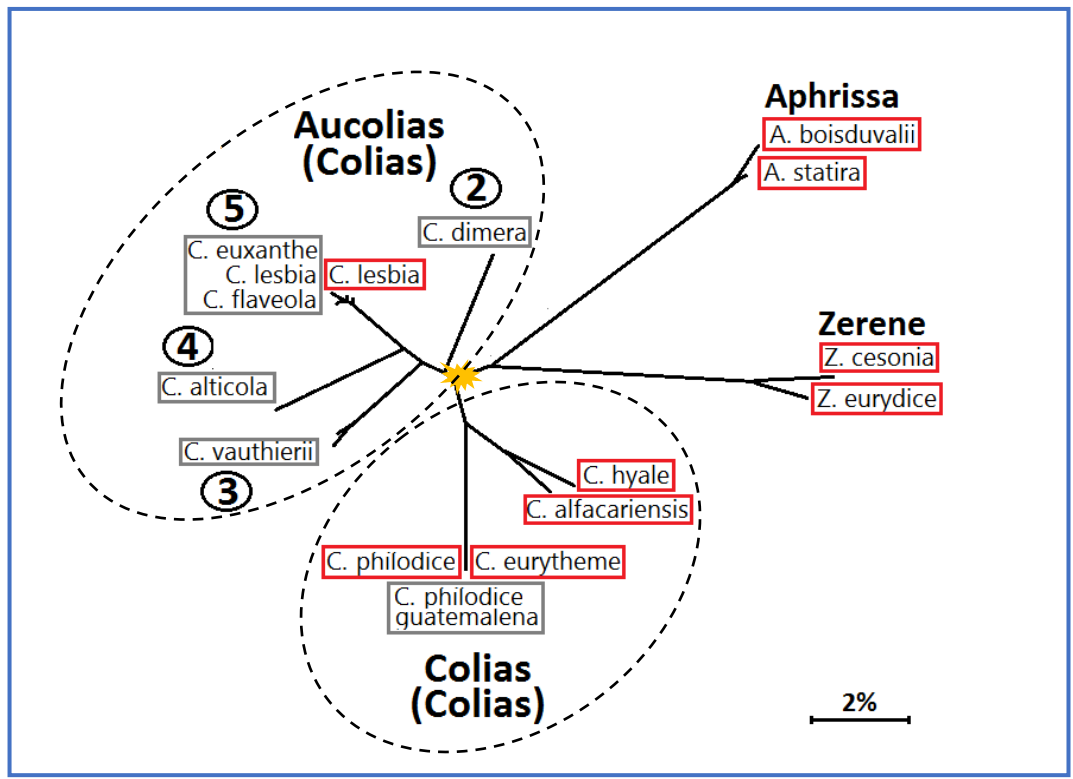

**Figure 9.** Unrooted phylogenic tree, generated applying 'barcoding' made on the basis of the representative data stemming from the SACOL project (refer to Fig. 6), combined with such ones stemming from GenBank, sourced from specimens of the few genetically closest *Colias sp.* outside S. America (*C. philodice*, *C. eurytheme*, *C. hyale*, and *C. alfacariensis*), and one more S. American specimen (*C. lesbia*). Also, the data of 'barcoding' of representatives of taxonomically closest genera (*Zerene*: sister to *Colias*) and *Aphrissa* (sister to *Zerene*) are added. The orange asterisk tentatively marks the 'bifurcation point', adherent to a process that might have led to separating subgenus *Aucolias* (*Colias*) of S. America from its counterpart (now sister) subgenus *Colias* (*Colias*).

Furthermore, for the new subgenus *Aucolias*, to which all S. American *Colias* belong, a type species must be designated, as required by the Code. According to the type of *Aucolias* branching (see Fig. 5), the most divergent *sp.* among the entries and so expectedly evolutionary oldest is *C. alticola*. Therefore, *C. alticola* is designated to be the type *sp.* of subgenus *Aucolias*.

Note that as there have been several subgenera proposed in the past, to which one or more of the *sp.* under consideration have been assigned (see Section 6), this would turn to be that subgenus *Aucolias* may become a junior synonym of the oldest among those (hereafter disregarded) subgenera, *Scalidoneura* BUTLER, 1871. However, given that designation of a new genus (subgenus) is not strictly regulated by the Code, I consider it merely reasonable to render the new subgenus name, *Aucolias*, as it stems from the new knowledge established here via molecular phylogeny means.

As seen from comparison of Figs. 8 and 9, the subgenus *Aucolias* is sister to the group that comprises the rest of worldwide *Colias*, which accordingly becomes the other *Colias* subgenus *s. str*. Note that, in accordance to the history of the *Colias* subgenera concept (see Section 6), the name *Colias* as subgenus *s. str.* should be preserved for the latter group of *sp.* (with the type *sp.* being *Colias hyale*). Note that *Colias* (*Colias*) – in contrast to *Aucolias* (*Colias*), composed of 6 *sp.* only, is highly diversified in internal splitting, eventually yielding >70 *Colias sp.* But, eventually, these two subgenera ought to have one, and the same, ancestor, thus revealing monophyly of the *Colias* genus *s. lat.*, gathering these two without 'cross-talk'. An event possibly underlying this splitting is schematized in Fig. 9 by the orange asterisk.

To the end of this section, the tabulation summarizing the data presented above in the form of phylogenic trees is given in Fig. 10, allowing one's overall view on their statistical parameters.

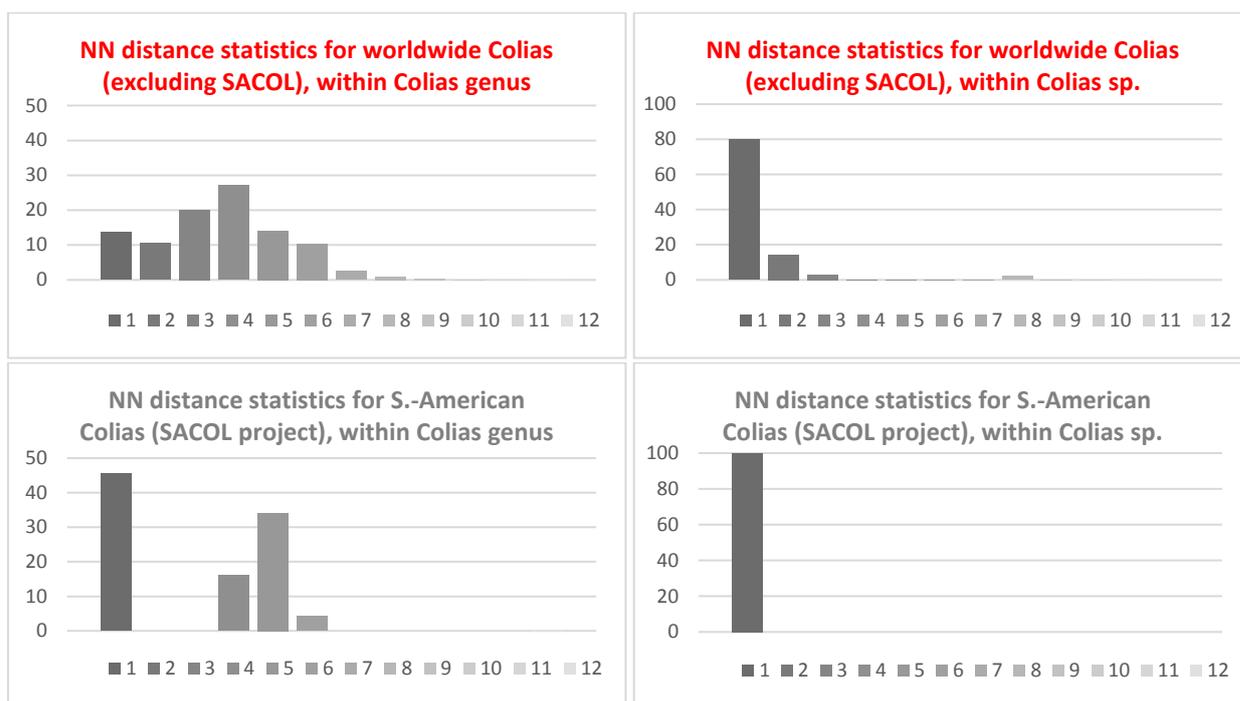

**Figure 10.** Histograms of NN distance statistics (as percentage of the total number of specimens covered by the analyses) for *Colias* sp. within the genus (on the left), and within each sp., treated separately (on the right). The two upper histograms are built on the data base from GenBank (excluding the ones from the SACOL project), and the two lower ones are built exclusively on the SACOL data base.

## 5. DISCUSSION

As known, the *Pieridae*, as a family, remain the most poorly investigated diurnal butterflies. The same unhappy circumstance, but with more 'internal' mess, is still preserved with *Colias* as genus and its subgenera [BRABY 2005], as well as with the relationship of these as a whole with systematically close genera, among which the closest one, as has been recently established, is *Zerene* [POLLOCK ET AL. 1998]. First unambiguous steps to clarify and fixate the situation with the higher than *sp.* level systematics of *Colias* have been recently attempted, employing contemporary molecular phylogenetics methods [POLLOCK ET AL. 1998, WHEAT ET AL. 2005, BRABY ET AL. 2006, WHEAT ET AL. 2008, LAIHO ET AL. 2013, WAHLBERG, ET AL. 2014, KRAMP ET AL. 2016]; however, none of these studies dealt with S. American *Colias*. The two exclusions are the recent researches [MURILLO-RAMOS ET AL. 2016, LAVINIA ET AL. 2017]: in the first work, the case of *C. dimera*, but sole among the other representatives of American *Coliadinae*, was considered from the metamorphological point of view, while in the second one, the position of *C. lesbia* was defined, employing barcoding, among the complex of diurnal butterflies of S. Argentina.

The present study is dedicated to examination of the complex of S. American *Colias* by means of molecular phylogeny for the first time to the best of my knowledges. As demonstrated above, the current, yet primary, insight has allowed uncovering interesting patterns of NN divergences among S. American *Colias* (treated as a separate concise complex), as well as between this complex and 'remaining' *Colias sp.*, distributed worldwide but off S. America.

As follows from the above reported data (see Figs. 5 to 9 and the notes therein), the complex comprising all S. American *Colias* is a monophyletic clade, never reported before. Its main essence is larger or much larger than in overseas *Colias* relative NN distances, or divergences, discriminating the entries (*sp.*). The internal structure of S. American *Colias* is characterized in terms of four clusters, with three of these being monospecific (defined and represented by *C. dimera*, *C. vauthierii*, and *C. alticola*), and the fourth gathering three weakly differentiating, or emerging, *sp.* (*C. lesbia*, *C. flaveola s. lat.*, and *C. euxanthe*). Besides, once this complex is straightforwardly characterized in terms of phylogenetic relations to the one comprising all other worldwide *Colias* (refer to Figs. 8 and 9), it becomes apparent that there is notable distinction between the two. Interestingly, S. American *Colias sp.* and overseas *Colias sp.* are never intermixed in the phylogeny trees. This has forced me to assign the clade formed by all S. American *Colias* (viz. *sp. C. dimera, C. vauthierii, C. alticola, C. lesbia, C. flaveola*, and *C. euxanthe*) as a new subgenus – *Aucolias* (*Colias*).

Certainly, a more detailed research would be required to seek for additional evidences in support of the proposed schematization of *Colias* phylogeny, say, trialing more versatile fragments of genome. However, it seems doubtful that the highlighted internal structure of S. American *Colias* as a delimited complex would become different in its key aspects.

As one can elucidate from the presented above data on ease and effectivity of hybridization among S. American *Colias*, (see Table 2 and the notes therein), this seem to be an impactful phenomenon for their evolutionary diversification.

Given the types of both adult and larval color patterns and morphology, two main evolutionary clades, or lineages (see Section 7) appear to have developed in S. America. One of them, represented by *C. lesbia*, is characterized by orange-yellow coloration, a solid black wing border in males, an androconial scale patch on the dorsal hindwing of males, and a mostly green color pattern in the larva. By contrast, the second lineage, represented by super-*sp. flaveola* (*C. mossi*, *C. flaveola*, and *C. weberbaueri*) is characterized by white to yellowish coloration, often with heavy melanic scaling, a broken black wing border in males, instability in development of androconial scale patch in males, and a striped color pattern in the larva. Besides, the overall type of wing morphology (especially in the hindwings) appears to be very similar in these two groups of S. American *Colias*. The said brings in a hypothesis that the extensive hybridization has taken place between the ancestors of lineages *lesbia* and *flaveola* in S. America over millions of years (Myr). Furthermore, *C. euxanthe* may be thought to have an entirely reticulate origin, resulting from a hybrid fusion between the two lineages.

One of the breaking news of the present study is that it brings in, at the basis of CO I mitochondrial barcoding, the evidences for a hybrid history of a big piece of S. American *Colias*. What is particularly interesting is that majority of S. American *Colias* (viz. *sp. C. mossi, C. flaveola s. lat., C. lesbia*, and *C. euxanthe*) share haplotypes, suggesting that these haplotypes have had a strong selective advantage. The acquisition of the dominating haplotypes (belonging to cluster '5' in Fig. 5) appears to be basically the result of hybridization. On the other hand, *C. dimera* (of northern S. America, reminiscing *sp.* of genus *Zerene* of the Americas) and *C. vauthierii* (of southern S. America, resembling *C. electo* LINNAEUS of Africa), constituting, correspondingly, clusters '2' and '3' in Fig. 5, have been apparently not involved in such-kind hybridization as evidenced by the barcoding data – probably because of long-term geographic separateness and so the absence of gene flow from outside. The situation with *C. alticola* that forms own lineage (or cluster '4' in Fig. 5) currently is not entirely clear. This taxon has usually been regarded in the past as *ssp.* of *C. euxanthe*, but the present study reveals its strong genetic divergence from the rest of S. American *Colias*. A tentative explanation for this effect might be that hybridization in the last case was not effective or even possible due to independent evolution of this *sp.* for long time, circumstanced by survival at extreme highlands in Ecuadorian altiplano (at >4000 m a.s.l.).

## 6. NOTES ON THE INTERNAL SYSTEMATICS OF COLIAS AS GENUS

Let me gain insight into the current (but prior to this study) *Colias*' systematics in terms of gradating the genus on subgenera.

In Table 4, a resume of how subdivision on subgenera of the *Colias* genus developed historically is snapshotted. Note that, in all attempts to deal with the issue, the authors dealt with phenomenological terms, viz. with the butterflies' phenotypes, genitalia's details, geographic distributions, etc. As seen from the Table, to the end of the 20th century, it was created a mess in the pattern; e.g., see the inconsistences highlighted (in bold) the different subgenera being created whereas based on same type taxa).

**Table 4:** Historical view on the development of the *Colias* subgenera concept. Highlighted are: in yellow – subgenera proposed for S. American *Colias*; in bold – differently named subgenera but based on same type entities.

| Subgenus | Type sp. | Range |
|---|---|---|
| *Colias* Fabricius, 1807 | **C. hyale** | Eurasia |
| *Eurymus* Horsfield, 1829 | **C. hyale** [in fact, *C. erate*] | Eurasia |
| *Scalidoneura* Butler, 1871 | *C. euxanthe* | S. America |
| *Eriocolias* Watson, 1895 | *C. edusa* [= *C. croceus*] | Europe and N. Africa |
| *Coliastes* Hemming, 1931 | **C. hyale** | Eurasia |
| *Mesocolias* Petersen, 1963 | *C. vauthierii* | S. America |
| *Protocolias* Petersen, 1963 | **C. imperialis** [= **C. ponteni**] | S. America or Pacific |
| *Palaeocolias* Berger, 1986 | **C. ponteni** [= **C. imperialis**] | S. America or Pacific |
| *Neocolias* Berger, 1986 | *C. erate* | Eurasia |
| *Eucolias* Berger, 1986 | *C. palaeno* | N. Eurasia and N. America |
| *Paracolias* Berger, 1986 | *C. dimera* | S. America |
| *Similicolias* Berger, 1986 | *C. lesbia* | S. America |
| *Asiocolias* Korb, 2005 | *C. christophi* | C. Asia |

Probably, this unsatisfactory situation forced L. Berger [Berger 1986] to revise the internal systematics of *Colias* as genus; this was a historically recent action to stabilize its subdivision, which – with higher or lower confidence – is utilized to-date. It is worthwhile noting that Berger's systematics was grounded on dealing with an only few morphological characters, sometimes weakly diagnosable in *Colias* butterflies, which led him to propose 6 (!) new subgenera for *Colias*, with some of these insufficiently supported. Berger's view on the matter has since then been criticized many times by experienced specialists. Recently, when the point of *Colias* genus' subdivision was revisited by M. Braby [Braby 2005], it has been stated that "Berger's subgenera", besides *Colias* (*Colias*), "*can only be listed tentatively until further studies*".

Therefore, because of mist dazed on *Colias* systematics (as shown above), on one hand, and, on the other hand, given that the data on the molecular phylogenetics of S. American *Colias* reported in Section 4 gave real news about their place in the *Colias* phylogeny tree, it is reasonable to propose here an alternative systematics for *Colias*.

The phylogram shown in Fig. 11 specifies the novel *Colias* subdivision into – as I believe – more supported subgenera and lineages (former subgenera) that are now the internal elements of the former; see the Table in the figure.

This scheme is built on main 'bricks' being typic *sp.* designating each of all *Colias* lineages (instead of former Berger's subgenera), whilst it accounts for the new knowledge received after barcoding S. American *Colias*. As seen, the last group is a necessary chain – unintentionally omitted or overlooked by researches in the past – in the revised structure of *Colias* as genus. Note that, where it was possible and non-contradicting with the laws established in the above made phylogenetic reconstruction, the primary subdivision elements within the genus are preserved in accord with the hierarchy (refer to Table 4), to maintain historical continuity of the terms.

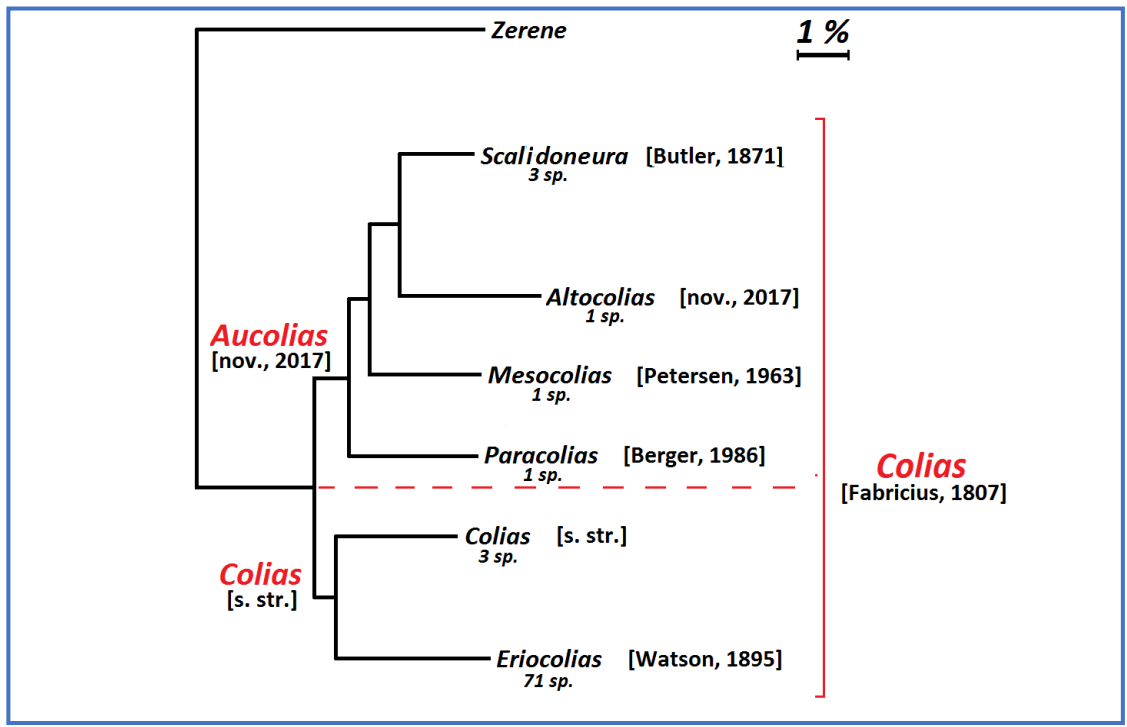

| Genus | Subgenera | Lineages (ex-subgenera) | Type sp. | Number of sp. |
|---|---|---|---|---|
| Colias FABRICIUS, 1807 | Colias s. str. | Colias s. str. | C. hyale | 3 |
| | | Eriocolias WATSON, 1895 | C. croceus | 71 |
| | Aucolias nov. | Paracolias BERGER, 1986 | C. dimera | 1 |
| | | Mesocolias PETERSEN, 1963 | C. vauthierii | 1 |
| | | Scalidoneura BUTLER, 1871 | C. euxanthe | 3 |
| | | Altocolias nov. | C. alticola | 1 |

**Figure 11.** Top: hierarchical phylogeny of *Colias* as genus (generalized scheme); bottom: tabulated list of *Colias*' higher-than-sp. classification (genus, subgenera, lineages, type sp., number of sp., and average divergences within each lineage).

Eventually, *Colias* genus is split in 2 subgenera: *Colias* (*Colias*) and *Aucolias* (*Colias*). The first of them includes, in total, 74 *Colias sp.* occurring in Eurasia, N. America, and Africa, while the second (new) one comprises 6 *Colias sp.* of S. America. Each of the new subgenera is composed of a few lineages (most of which recall some of the former subgenera): for *Colias* (*Colias*), these are *Colias* (3 sp. – *C. hyale*, *C. alta* STAUDINGER, *C. alfacariensis*) and *Eriocolias* (71 sp. – all other Eurasian, N. American, and African *Colias*); for *Aucolias* (*Colias*), they are *Paracolias* (1 sp. – *C. dimera*), *Mesocolias* (1 sp. – *C. vauthierii*), *Scalidoneura* (3 sp. – *C. euxanthe*, *C. lesbia*, *C. flaveola s. lat.*), and new *Altocolias* (1 sp. – *C. alticola*).

## 7. CALLS FOR FUTURE WORK

Supposedly, the proposed version of the higher than *sp.* level *Colias* systematics may instill motivation for further studies in the field by interested people, especially by means of molecular phylogeny methods.

First, a quite interesting point regarding S. American *Colias* is the relationships between numerous 'white-to-black' (in dorsal coloration in males) entities, gathered in the present study under the name *C. flaveola s. lat.*: refer to Table 1 and Fig. 7 (right-hand panel in bottom). The group of these taxa is hard to systemize beyond doubt. Some authors treat all these as *ssp.* of one, and the same, *sp.* ('super-*sp.*'), as [LAMAS 2004] but others consider all them as separate *sp.*, as [VERHULST 2000 a-b, VERHULST 2013, GRIESCHUBER ET AL. 2012]. Although there are evidences in support of such internal systematics (quite different phenotypes, behaviors, environmental conditions, host plants, etc., inherent to each taxon), a final decision on real relations between them is a matter of future studies (e.g. by molecular phylogeny methods). The current work, where barcoding of a single mitochondrial fragment of CO I subunit has been employed, unfortunately revealed incapability to resolve the problem: S. American 'black-to-white' *Colias* of S. America are indistinguishable after simple barcoding. Thus, a more diversified phylogenic study is required to properly address the point.

Second, the most intriguing issue is a position in the phylogeny tree for *Colias* of the legendary *C. imperialis*. (Note that this *sp.* is not listed in Fig. 11 to avoid any confusing misinterpretation.) This butterfly – being the most primitive *sp.* among *Colias* on all sides – may be a key element for understanding the true hierarchy and evolutionary history of *Colias* as genus. In this sense, a question to pose is: whether this butterfly presents a 'lost' chain in between genera *Zerene* and *Colias*, or exemplifies a kind of 'archipelago' *sp.* that has received its primitive appearance because of long-time occurrence in a non-alternating by gene flow ambience in the absence of neighboring relatives?

Third, the point that deserves attention is: why – despite the burst of speciation happened in the past (while relatively recent) with *sp.* of subgenus *Colias* (*Colias*) of the northern hemisphere (refer to Figs. 8 and 11) – none reminiscing that happened with subgenus *Aucolias* (*Colias*) of S. America? Can explanations be global periods of glaciation and fading, sustained in Eurasia and N. America (which might lead to such types of inter-continental gene migrating and eventual speciation via, say, the trans-Beringia scenario [ESTOUP AT AL. 2010, VILA ET AL. 2012]), or appearance of local hot spots of survival-in-refuge with posterior invasions and reinvasions into nearby areas at climate's cycling? Both kinds of processes can be arguable in the attempt to address the speciation boom in the past and, consequently, the current overall diversity of subgenus *Colias* (*Colias*) of the northern hemisphere. However, the only second explanation is likely to impact, but with a limited effect, the own evolution of subgenus *Aucolias* (*Colias*) of S. America, which probably evolved 'autonomously' after splitting from a common with *Colias* (*Colias*) ancestor. The mentioned controversies might underline speciation growth in the first case (in *Colias* of the northern hemisphere) and its lower effect in the second one (in *Colias* of the southern hemisphere). It cannot be ruled out that the mentioned trans-Beringia scenario might have been gone, in the case of *Colias* evolution, 'forward' (from America to Eurasia) and 'backward' (from Eurasia to America), with the first move being a kind of the Camelid scenario. As was argued by SHAPIRO [SHAPIRO ET AL. 2007], the high-altitude *Pierini* of S. America and Eurasia are not each other's close relatives but – as seen from the phylogeny data presented above (refer to Figs. 8, 9, and 11, e.g.) – this such type of scenario may apply to *Colias*. Besides, an ancestor of sister genera *Colias* and *Zerene* [POLLOCK ET AL. 1998] lived, most probably, in the Americas, as no *Zerene* are found to the North of California; furthermore, this 'law' may hold for a common ancestor of all *Coliadinae* [BRABY ET AL. 2006]. A process of splitting of *Colias* and *Zerene* might have happened in Central or S. America in a few Myr back. Indeed, NN distance between the genera is ~11% (Figs. 9 and 11), corresponding to ~8 Myr if one uses the estimate for the rate of mutations in mitochondrial CO I part of genome, ~1.5% per 1 Myr [BRABY ET AL. 2006]. In turn, splitting of subgenera *Aucolias* (*Colias*) and *Colias* (*Colias*) might have happened ~5 Myr ago (NN distance between the two subgenera is ~7%). Accordingly, speciation startup within *Aucolias* of S. America might have been brought in ~3 Myr, i.e. at least by ~1 Myr earlier than that within *Colias* (*Colias*) of the northern hemisphere. The latter estimates are relevant as to the type of branching in the common phylogeny tree (Figs. 8 and 11), as to the data summarized in Fig. 10: as seen, gravity center of the histogram of NN distances for the members of subgenus *Aucolias* (but excluding emerging and genetically indistinguishable *sp. C. lesbia*, *C. euxanthe*, and *C. flaveola*) is ~5%, a bigger value than that (<4%) for subgenus *Colias s.str*.

Fourth, a recent study on European *Colias palaeno* [KRAMP ET AL. 2016] should be referred to, where a more than 9% divergence between CO I haplotypes within a set of populations has been established, i.e. much higher genetic diversification within this *sp.* than within all other known *sp.* of the genus. Interestingly, this result reminds an earlier study on molecular phylogeny of a group of 'orange' N. American *Colias hecla / Colias canadensis* [POLLOCK ET AL. 1998], for which – given significant intraspecific genetic divergences between the geographically and habitually different entries analyzed – the authors revealed possible occurrence of cryptic *sp.* inside this group of *Colias*. These results seem to be important for future phylogenetic reconstruction of the genus *Colias*.

**ACKNOWLEDGMENTS**

The author is gratefully thankful to EVGENY ZAKHAROV (Canada), ARTHUR SHAPIRO (USA), WARD WATT (USA), ROBERT WORTHY (UK), and ROBERT BORTH (USA) for valuable comments and kind assistance concerning this research.